\documentclass[11pt,a4paper]{article}
\usepackage{jcappub}
\usepackage{booktabs}
\usepackage{pgfplotstable}
\usepackage{graphicx,graphics,amssymb,amstext,amsmath,bbold,float}
\usepackage{epstopdf} 

\definecolor{nblue}{RGB}{0 102 204}

 \title{The shape dependence of chameleon screening}
\author[a]{Clare Burrage,}
\author[a]{Edmund J. Copeland}
\author[a]{Adam Moss}
\author[a]{and James A. Stevenson}
\affiliation[a]{School of Physics and Astronomy, University of Nottingham, Nottingham, NG7 2RD, United Kingdom}
\emailAdd{clare.burrage@nottingham.ac.uk}
\emailAdd{ed.copeland@nottingham.ac.uk}
\emailAdd{adam.moss@nottingham.ac.uk}
\emailAdd{james.stevenson@nottingham.ac.uk}
\abstract{Chameleon scalar fields can screen their associated fifth forces from detection by changing their mass with the local density. These models are an archetypal example of a screening mechanism, and have become an important target for both cosmological surveys and terrestrial experiments. In particular there has been much recent interest in searching for chameleon fifth forces in the laboratory.  It is known that the chameleon force is less screened around non-spherical sources, but only the field profiles around a few simple shapes are known analytically. In this work we introduce a numerical code that solves for the chameleon field around arbitrary shapes with azimuthal symmetry placed in a spherical vacuum chamber. \textcolor{blue}{ We find that deviations from spherical symmetry can increase the chameleon acceleration experienced by a test particle, and that the least screened objects are those which minimize some internal dimension. For the shapes considered in this work, keeping the mass, density and background environment fixed,  the accelerations due to the source varied by a factor of $\sim 3$.} }
\begin{document}
\maketitle

\section{Introduction}

Is there a light scalar particle mediating a long range fifth force in our universe? Despite a century of research, we still do not have a conclusive answer to this question.  We know that a canonical scalar, with a small mass is in conflict with astrophysical and terrestrial tests of gravity unless it is extremely weakly coupled \cite{Adelberger:2003zx}.  We also know, however, that allowing the scalar to have non-trivial self interactions leads to a much more varied phenomenology, which avoids the experimental constraints through a variety of mechanisms collectively  known as screening \cite{Joyce:2014kja}.

The chameleon model \cite{Khoury:2003aq,Khoury:2003rn} is one of the most commonly studied models of screening, and is an important test case for understanding how such scalar fields behave and how they can be constrained experimentally. The chameleon scalar has been linked to models of dark energy, and may modify  structure formation on the very largest scales in the universe \cite{Brax:2005ew,Brax:2013mua}. With non-trivial self interactions in its potential it has a mass which depends on the density of the local environment.  This gives rise to the {\it thin shell mechanism} whereby on any given background sufficiently small and diffuse  objects may source and experience strong fifth forces, but sufficiently large and dense  objects will be screened from the fifth force.\footnote{These conditions will be made precise in the next section.} This gives rise to an emergent violation of the weak equivalence principle. 

The fact that not all objects are screened from the chameleon fifth force offers opportunities for experimental searches, if forces can be measured precisely on sufficiently small objects. It has recently been realized that in laboratory vacuua, atomic nuclei \cite{Clare_ProbingDark,Hamilton:2015zga,Elder:2016yxm}, neutrons \cite{Brax:2011hb,Ivanov:2012cb,Brax:2013cfa,Jenke:2014yel,Pignol:2015bqa,Lemmel:2015kwa,Li:2016tux}, and silicon microspheres \cite{Rider:2016xaq} all satisfy this condition, in at least some part of the chameleon parameter space.  This has lead to impressive new constraints on the parameter space, and the prospect of either directly detecting a chameleon force, or ruling out the model completely in the near future. 

No experiment yet has the sensitivity to detect the chameleon fifth force between two such small objects, and so the current approach is to study the attraction between one macroscopic object which is screened, and one microscopic object which may not be. Changing the shape of the macroscopic source, has been shown to change the amount of screening, and has the potential to increase the magnitude of the chameleon fifth force \cite{Burrage:2014daa}. However the study of the shape dependence of chameleon screening has so far been limited to simple shapes which can be treated analytically; for example ellipsoids are less screened than spheres of the same mass \cite{Burrage:2014daa}.

In this work we will extend this study numerically, looking at how different shapes change the amount of chameleon screening and attempt to determine the optimal source shape for a chameleon experiment.  We focus on the chameleon model, but the numerical code developed in this work, can be extended to study other theories with screening, where the amount of screening has also been shown to depend sensitively on the source shape.

In the following section we introduce the chameleon model, and review how the screening mechanism works. In Section \ref{sec:num} we describe the numerical code, which used the finite element method with mesh refinement, to  solve for the chameleon profile around  different source shapes within a spherical vacuum chamber. In Section \ref{sec:shape} we determine how the screening is affected by changing the source shape by solving for four characteristic source shapes - sphere, ellipsoid, torus and cardioid.  Finally, in Section \ref{ShapeOptimization}  we consider arbitrary shaped sources by using Legendre polynomials as a basis to construct the shape of the surface, and by going to fourth order in these polynomials we determine which shape maximises the chameleon acceleration.   We conclude in Section \ref{sec:conc}.

\section{The Chameleon}
The chameleon is a theory of a non-minimally coupled scalar field, $\phi$.  In the Einstein frame its action is
\begin{equation}
S= \int d^4 \sqrt{-g} \left[M_P^2R-\frac{1}{2}g^{\mu\nu}\partial_{\mu}\phi\partial_{\nu}\phi -V(\phi)\right] + S_m(\psi_i,A(\phi)g_{\mu\nu})
\end{equation}
where $g_{\mu\nu}$ is the Einstein frame metric, $R$ the associated Ricci scalar, $M_P$ the reduced Planck mass,  $V(\phi)$ is the chameleon potential, and $S_m$ is the action for matter fields which we generically denote as $\psi_i$.  These matter fields move on a metric which is a rescaling of the Einstein frame metric with a function of the chameleon field $\tilde{g}_{\mu\nu}=A(\phi)g_{\mu\nu}$.  A particular realization of the chameleon model requires specifying $A(\phi)$ and $V(\phi)$. 

The resulting chameleon equation of motion is 
\begin{equation}
\Box \phi= \frac{d V}{d\phi}+\frac{1}{M}T_{\mu}^{\mu}
\label{eq:chameom}
\end{equation}
where we have approximated the coupling function by the  two lowest order terms in its Taylor series, $A(\phi)= 1 +(\phi/M) +\mathcal{O}(\phi/M)^2$, so that $M$ is an energy scale which controls the coupling to matter.  It can be checked that higher order terms in this expansion remain small in the simulations which we discuss here.  $T_{\mu\nu}$ is the energy momentum tensor of the matter fields, and the chameleon is sourced only by the trace $T^{\mu}_{\mu}$. Note that this means there is no direct coupling between the chameleon and photons.  

For this scalar field theory to be a chameleon model we require that there is a finite value of $\phi$ for which the right hand side of equation (\ref{eq:chameom}) vanishes, and that this point be a minimum of the effective potential 
\begin{equation}
V_{\rm eff}(\phi)= V(\phi) +\frac{\phi}{M}T_{\mu}^{\mu}
\end{equation}
for non-relativistic matter distributions where $T_{\mu}^{\mu}>0$.  Our final requirement is that the mass of small fluctuations about this minimum be a function of $T_{\mu}^{\mu}$. These conditions are not particularly restrictive, if we consider polynomial potentials of the form $V(\phi)= \Lambda^4(\Lambda/\phi)^n$, with integer $n$, then we have a chameleon model if $n>0$, or if $n$ is an even negative integer strictly less than $-2$. 
The scalar potential with $n=1$ is a commonly used benchmark model in the literature, and in what follows we will specialize to this particular case for simplicity when presenting our results. However the numerical code developed in this work can be used for any choice of chameleon potential. 

In any given environment the chameleon wants to sit at the minimum of its effective potential.  In non-relativistic environments, where $T_{\mu}^{\mu}= \rho$, the local energy density of matter, for our choice of potential this field value is given by
\begin{equation}
\phi_{\rm min}(\rho)= \left(\frac{M \Lambda^5}{\rho}\right)^{1/2}
\end{equation}
and the mass of small fluctuations around the minimum is
\begin{equation}
m_{\rm min}^2(\rho) = 2\left(\frac{\rho^3}{M^3 \Lambda^5}\right)^{1/2}
\end{equation}
We can see directly that the mass increases as the local density increases. 

In environments where the local density varies the chameleon may not always be able to reach the value which minimizes its potential.  A simple example of this is a spherical vacuum chamber, where the chameleon does sit in the minimum of its potential in the walls.  In the interior of the vacuum chamber the field wants to grow, but may not have enough space to reach $\phi_{\rm min}(\rho_{\rm vac})$.  In this case the maximum value that the field reaches will be the one where the Compton wavelength of the chameleon is of the order of the size of the vacuum chamber $L$, so that $\phi_{\rm central} \sim (L^2\Lambda^5)^{1/3}$ \cite{Clare_ProbingDark}. 

A second example where the chameleon does not reach the minimum of its effective potential, is for a small, or diffuse object in a lower density background. If the object only causes a small perturbation of the chameleon field about its minimum in the background environment then it may not reach the value which minimizes the potential in the interior of the object.  For a spherical object of constant density, with mass $M_A$ and radius $R_A$ the field profile is 
\begin{equation}
\phi = \phi_{\rm bg}-\frac{1}{8 \pi R_A} \frac{M_A}{M}\left\{\begin{array}{lc}
3 - \frac{r^2}{R_A^2} & r< R_A \\
\frac{2R_A}{r}e^{-m_{\rm bg}r} & r> R_A
\end{array}\right.
\end{equation}
where $\phi_{\rm bg}=\phi_{\rm min}(\rho_{\rm bg})$ is the value of the field which minimizes the potential in the background environment, and $m_{\rm bg}=m_{\rm min}(\rho_{\rm bg})$ is the corresponding mass. This solution is valid when $(M_A/4 \pi R_A) \ll M \phi_{\rm bg}$.

If the size or the density of the object is increased the field profile leaves this weakly perturbing regime, and the field does reach the value which minimizes its effective potential at the center of the sphere. In this case the field profile is
\begin{equation}
\phi = \left\{\begin{array}{lc}
\phi_{\rm in} & r<S\\
\phi_{\rm in}+\frac{1}{8 \pi R_A} \frac{M_A}{M}\frac{r^3 -3S^2 r+2S^3}{r R_A^2} & S<r< R_A \\
\phi_{\rm bg}-\frac{1}{4 \pi R_A} \frac{M_A}{M}\left( 1-\frac{S^3}{R_A^3}\right)\frac{R_A}{r}e^{-m_{\rm bg}r} & r> R_A
\end{array}\right. \label{eq:thinshell}
\end{equation}
where $\phi_{\rm in}=\phi_{\rm min}(\rho_{\rm A})$, and $S$ is known as the thin shell radius which is fixed by 
\begin{equation}
S=R_A \sqrt{1- \frac{8 \pi}{3}\frac{M}{M_A}R_A \phi_{\rm bg}}
\end{equation}
In both of these cases we have assumed $m_{\rm bg}R_A \ll 1$, and $\phi_{\rm bg} \gg \phi_{\rm in}$.

The force that a test particle experiences when moving in these potentials is given by $\vec{F} =-\vec{\nabla} \phi/M$.  In this way we can see that as the chameleon is pushed into the non-linear regime as in Equation \ref{eq:thinshell}  the fifth force is suppressed as $S$ becomes close to $R_A$. This is known as the thin shell effect. 

The expressions for the chameleon profiles around non-spherically symmetric sources become rapidly more complicated.  One of the few other situations for which an analytic solution is known is when the source is an ellipsoid \cite{Burrage:2014daa}.  We do not reproduce the full form of the scalar field profile here, but we found that in the exterior of the ellipsoidal source the ratio of the chameleon to gravitational force is
\begin{equation}
\frac{F_{\phi}}{F_G}= 2\left(\frac{M_P}{M}\right)^2\left(1 - \frac{\xi_{\rm core}(\xi^2_{\rm core} - 1)}{\xi_0(\xi^2_0 - 1)}\right)
\end{equation}
where $\xi_0$ is the ellipsoidal `radius' of the source in spherical prolate coordinates, and $\xi_{\rm core}$ is the equivalent thin shell radius.  As the source becomes more ellipsoidal, keeping the total mass and density fixed, the screening becomes less efficient, and the fifth force less suppressed. 

We hypothesize that the weaker screening in the ellipsoidal case is because there are now two length scales relevant to describing the shape, a minimum and a maximum diameter.  Along the axis of the minimum diameter the scalar field has a much more limited amount of space within which to evolve.  This makes it much harder for the field to reach the minimum of its effective potential at the center.  It is this hypothesis that we wish to test in this work using numerical simulations.

\section{Numerical Approach}
\label{sec:num}
Theories of screening are intrinsically non-linear, which makes solving for the field profiles numerically a challenging problem.  Previous work has focused on simulating the effect of the chameleon on the formation of large scale structure in the universe \cite{Lombriser:2013eza,Brax:2013mua,Bose:2016wms}, and on simulations for specific atom interferometry experiments \cite{Elder:2016yxm,Schlogel:2015uea}.
In this work we will use the finite element method to solve these equations. We find that this is more suited to the highly non-linear problem we wish to tackle than the more traditional finite difference method, which struggles to resolve significant field variations over very short scales,  particularly on an equispaced grid. 
The finite element method  uses an  integral form of the field equations, which does not rely on a specific discretisation of the spatial coordinates.
It can also easily handle discontinuities in the source, which in our case are discontinuities in density between the interior and exterior of the source mass. 
Finally Neumann boundary conditions appear naturally within the formulation, and so do not have to be enforced by hand.
To solve the equations in this form, our numerical code relies on  the differential equation solver developed by the FEniCS Project \cite{FenicsBook}, launched in 2003, which has led to a state of the art collection of software libraries built around the finite element framework for both C++ and Python.

In what follows we will focus on the case of the chameleon model.
However the numerical code we have developed has been designed in such a way that the scalar potential and coupling function can be easily changed. This means that comparable data for other models of screening, such as the symmetron, can be easily obtained.

\subsection{Piecewise Linear Approximations and Finite Element}

\begin{figure}
\centering
\includegraphics[scale = 1, trim = 0cm 25cm 0cm 0cm]{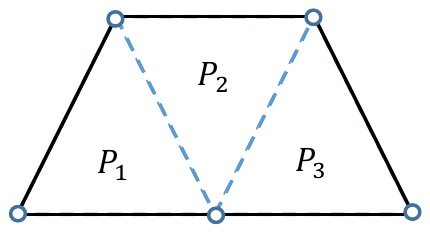}
\caption{Example triangulation of a trapezium. The polygon is broken into three triangular sub-domains $P_i$. Common edges are represented by the dashed blue lines to emphasize the continuity of a piecewise linear function across the boundary. The vertices used to calculate the unknown coefficients of each linear function are identified using circles.}
\label{triangulation}
\end{figure}

In this section we will briefly introduce the formalism necessary to describe the solution to a differential equation in terms of finite elements.
The central idea underlying this approach is that functions can be approximated, in a piecewise manner, by polynomials \cite{PiecewiseLinear}. 
To solve a set of equations over a two dimensional domain $\Omega_D$ we subdivide the domain into  a network of triangles, an example of which is shown in Figure \ref{triangulation}.  The index $i$ will label the vertices of the triangles, and the coordinates, $x_1,x_2 \in [0,1]$, describe the interior of each triangle.  The polynomials typically used are the set of linear functions $\mathbb{P}_1$ of the two variables, $x_1$ and $x_2$, such that
\begin{equation}
\phi(x_1,x_2) = a_0 + a_1 x_1 + a_2 x_2
\end{equation} 
 Within each (non-degenerate) triangle $P_i$, the unknown coefficients $a_0$, $a_1$ and $a_2$ of a function $\phi$ will be uniquely determined by the values of the function  at the vertices.  Furthermore, as the values of $\phi|P_i \in \mathbb{P}_1$ along an edge depend only on the values at the connecting vertices, the piecewise description of $\phi$ is automatically continuous across edges.

Denoting the $N$ vertices within the triangulation as $\textbf{p}_i$ where $ i = 1, 2, \dots, N$, it is useful to define a set of basis functions $e_i(x_1,x_2)$, which are `tent functions',  such that $e_i(p_j) = \delta_{ij}$; the piecewise continuous function $e_i$ is one at vertex $i$ and zero at all other vertices. Piecewise continuous functions $\phi$ over the domain $\Omega_D$ can therefore be built by taking linear combinations of these basis functions. 
\begin{equation}
\phi = \sum_{i = 1}^N \phi\left(\textbf{p}_i\right)e_i
\end{equation}
with constant coefficients $\phi\left(\textbf{p}_i\right)$.  

We have now broken the  space into finite (triangular) elements and used a local basis to construct a piecewise linear yet globally continuous function. This globally continuous function will be an approximation to the exact solution of a partial differential equation. 
Solving such a differential equation does not require a particular structure of the grid, instead the user is free to define the triangulation, on whichever domain most suits the problem. 

\subsection{Solving for the Chameleon Field within a Spherical Vacuum Chamber}

We will focus our attention on solving for the chameleon field profile around a static source in a spherical vacuum chamber. The equations of motion, to be written in finite element form are:
\begin{equation}\label{fieldEquation}
\begin{split}
\nabla^2\phi &= -\frac{\Lambda^5}{\phi^2} + \frac{\rho}{M}, \hspace{1cm} \rm{in} \>\Omega \\ 
\phi &= \phi_{\rm{min}}(\rho_{\rm wall}), \hspace{1cm} \rm{on} \> \partial\Omega, \\
\end{split}
\end{equation}
where $\Omega$ is the domain, which we take to be a cube in 3-dimensions and a square in 2-dimensions,   and  $\partial\Omega$ is its boundary.

We are interested in the behavior of the chameleon within a spherical vacuum chamber. Assuming that there is at least azimuthal symmetry, we study the idealized case of a spherical vacuum chamber, which reduces to a circle within the square domain. 
The width of the simulation box will be denoted $w$ and the radius of the vacuum chamber will be given by $R_C<w/2$. For simplicity we will assume that all points within the domain $\Omega$ at a radial distance from the center $r>R_c$ are within the wall, and so the density there is $\rho=\rho_{\rm wall}$. 

To solve this equation using a finite element solver, the equations of motion must be put into an integral form. Using Green's theorem we find:
\begin{equation}\label{weakForm}
\int_{\Omega} \left(\frac{-\Lambda^5}{\phi^2} + \frac{\rho}{M}\right)\varphi + \int_{\Omega}\nabla\phi\cdot\nabla\varphi = \int_{\partial\Omega}\left(\partial_n\phi\right)\varphi
\end{equation}
where $\phi$ is  the  chameleon field and $\varphi$ is a test function. The term appearing on the right hand side $\partial_n\phi$ denotes the exterior normal derivative,  $\nabla\phi\cdot\hat{\textbf{n}}$. The unit normal vector $\hat{\textbf{n}}$ is defined for points along the boundary $\partial\Omega$, pointing outwards from the domain $\Omega$.

A common choice for the test function, $\varphi$, is to use   the basis functions $\sum e_i$ used to build the finite element space,  known as the Galerkin method.  This  has the advantage that the resulting matrices to be handled by the solver are symmetric, and relaxes the smoothness requirements on the field $\phi$ \cite{PiecewiseLinear,Imperial}. Further, as this basis is almost orthogonal 
the matrices handled by the solver are sparse.

We  assume that the thickness of the walls of the vacuum chamber is always greater than the Compton wavelength of the field $\lambda_{\phi}^{(\rm{wall})}$ within the wall. This ensures that the field reaches $\phi_{\rm min}(\rho_{\rm wall})$  within the walls, at which point the Neumann boundary term $\partial_n\phi$ vanishes. If we define the half-width of the simulation mesh as $W$ this is consistent when
\begin{equation}
W - R_C > \frac{1}{\sqrt{2}}\left(\frac{\Lambda^5 M^3}{\rho^3_{\rm{wall}}}\right)^{1/4} 
\end{equation} 
For example, for $\Lambda = 10^{-10}\mbox{ GeV}$, $M = M_{\rm{PL}}$ and $\rho_{\rm{wall}} = 1\>$g/cm$^{3}$, this corresponds to $W - R_C > 0.1\rm{ mm}$. In what follows we take the radius of the vacuum chamber to be $15\rm{ cm}$ and the system half-width as $18\rm{ cm}$ which satisfies this inequality for all parameter choices of interest. In this parameter space  the point at which $\phi=\phi_{\rm min}(\rho_{\rm wall})$  always lies  close to the surface of the walls and so we approximate the solution by  imposing Neumann boundary conditions at the surface of the wall.

Equation \eqref{weakForm} now becomes
\begin{equation}\label{penultiWeakForm}
\int_{\Omega} \left(\frac{-\Lambda^5}{\phi^2} + \frac{\rho}{M}\right)\varphi + \int_{\Omega}\nabla\phi\cdot\nabla\varphi = 0 \hspace{1 cm} \forall \varphi
\end{equation}
This is a semi-linear form, being non-linear in the unknown $\phi$ and linear in the test function $\varphi$. In order to call the finite element solver for this problem, we need to arrive at linear and bilinear forms for the fields $\phi$ and $\varphi$. To do this we linearize the inverse power term for $\phi$ around some input solution denoted $\phi_k$.
\begin{equation}
\frac{1}{\phi^2} = \frac{1}{\phi_k^2} - \frac{2}{\phi_k^3}\left(\phi - \phi_k\right) +\mathcal{O}(\phi-\phi_k)^2 = \frac{3}{\phi_k^2} - \frac{2}{\phi_k^3}\phi +\mathcal{O}(\phi-\phi_k)^2 
\end{equation}
So the chameleon equation in its integral form becomes
\begin{equation}\label{finiteElementForm}
\int_{\Omega}\left\{\nabla\phi\cdot\nabla\varphi + \left(\frac{2\Lambda^5}{\phi_k^3}\phi\right)\varphi\right\} = \int_{\Omega}\left\{\left(\frac{3\Lambda^5}{\phi_k^2} - \frac{\rho}{M}\right)\varphi\right\}
\end{equation}
The equations derived so far describes the full three dimensional system. We will now restrict ourselves to systems that are rotationally symmetric around one axis, and project the problem onto a two-dimensional space.  In practice we work with a dimensionless form of these equations, which we write explicitly in Appendix \ref{app:dim}.

\begin{figure}
\begin{minipage}[c]{0.329\textwidth}
\includegraphics[width = \linewidth]{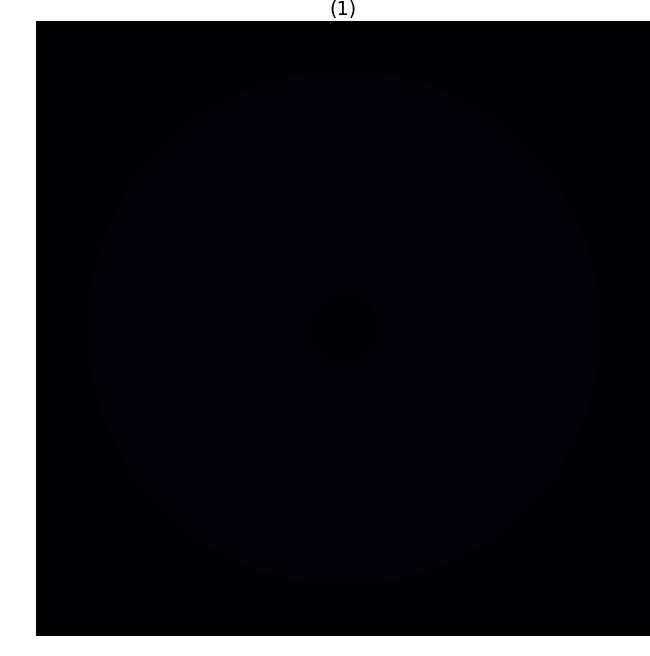}
\end{minipage}
\begin{minipage}[c]{0.329\textwidth}
\includegraphics[width = \linewidth]{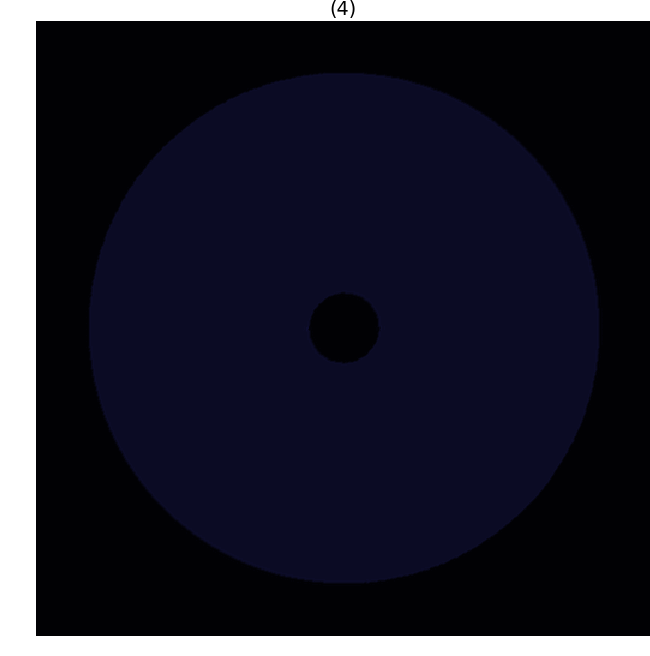}
\end{minipage}
\begin{minipage}[c]{0.329\textwidth}
\includegraphics[width = \linewidth]{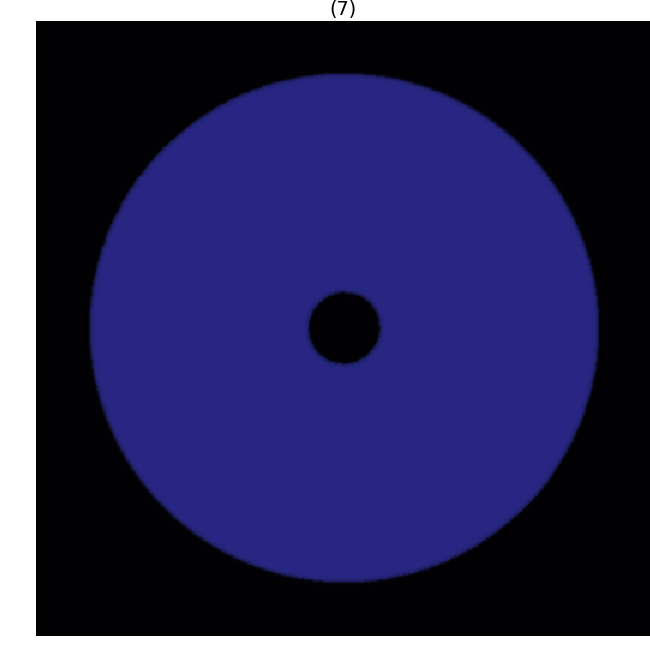}
\end{minipage}
\vskip\baselineskip
\begin{minipage}[c]{0.329\textwidth}
\includegraphics[width = \linewidth]{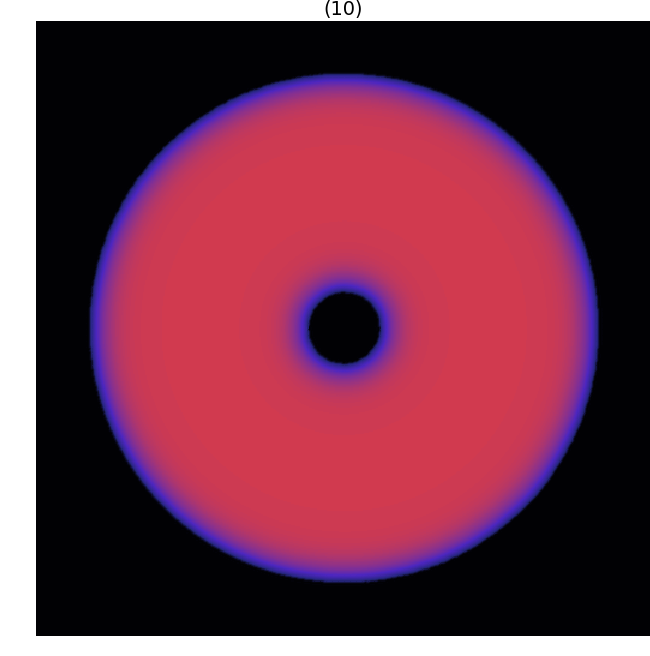}
\end{minipage}
\begin{minipage}[c]{0.329\textwidth}
\includegraphics[width = \linewidth]{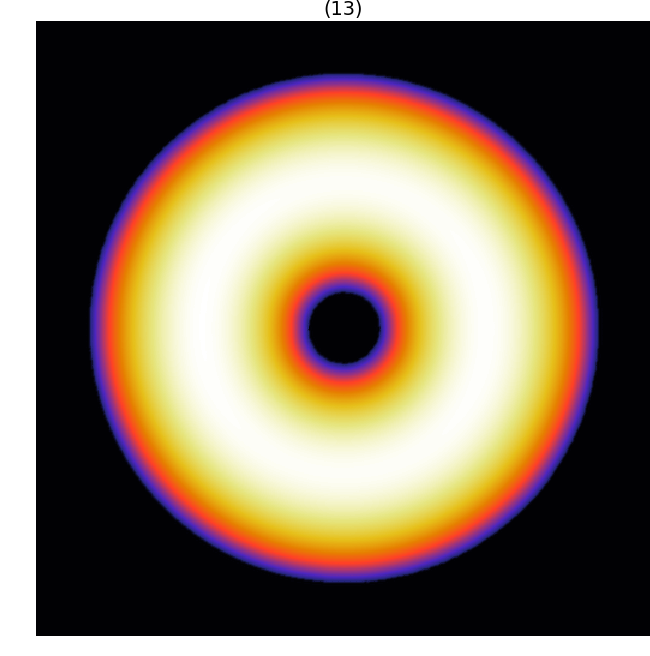}
\end{minipage}
\begin{minipage}[c]{0.329\textwidth}
\includegraphics[width = \linewidth]{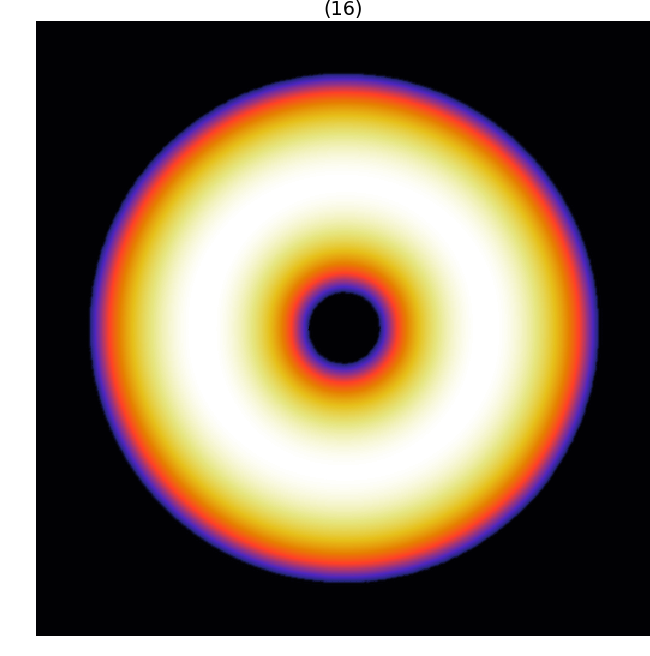}
\end{minipage}
\caption{Figure capturing the convergence of the chameleon field for a spherical source within a vacuum chamber over 16 iterations, six of which have been displayed here with the iteration number specified at the top of each snapshot. The colour scheme here is normalized across figures, where the field moves between the source expectation value (black) and the chamber expectation value (white). The coloring in this case following a linear interpolation scheme, from black through blue red and yellow to white.}\label{iterativeSolver}
\end{figure}

Calling a linear solver for the above problem involves recursively solving  \eqref{finiteElementForm} over a series of iterations. This strategy, referred to as Picard iteration, or the method of successive substitutions, uses the solution $\phi_k$ to the linear problem from iteration $k$ as the basis to search for an improved solution for iteration $\phi_{k+1}$. Figure \ref{iterativeSolver} shows how this  tends towards a converged solution  for a spherical source object within a spherical vacuum chamber after $16$ iterations. The field was chosen to take the value $\phi_{\rm{min}}^{\rm{(wall)}}$ everywhere in the   domain for the first iteration.

We use the vector norm  $\left|\phi - \phi_k\right|$  to estimate the errors for each iteration. This  can be shown to relate to the residual $r = L(\varphi) - a(\phi_k, \varphi)$ where $L(\varphi)$ and $a(\phi_k, \varphi)$ are the semi-linear and bilinear forms respectively,  appearing in \eqref{finiteElementForm}. Typically objects with stronger curvatures require a greater number of iterations in order to reach convergence. We required the residuals to be smaller than $10^{-6}$, to say that the solution had converged, and set an upper bound of  $30$ iterations before terminating the solver.

\section{Comparison of Fifth Forces for Different Shaped Sources}
\label{sec:shape}

In this section we explore whether breaking the spherical symmetry of the source mass would increase the likelihood that a chameleon force could be detected in a laboratory experiment. We wish to determine which source shape gives the largest chameleon force, and whether this is universally the best choice, or a function of the chameleon parameters. 
The finite element solver was called for a number of source geometries: sphere, ellipsoid, torus, and cardioid. Consideration of the spherical and ellipsoidal results was primarily to verify the outcomes of the numerical simulation against the known analytic solutions of \cite{Clare_ProbingDark} and \cite{Burrage:2014daa}, in addition to providing reference sources for the comparison of the other geometries. The toroidal source was taken as it is thought to offer more desirable systematics in the context of an atom interferometry experiment. The final source geometry under consideration was the cardioid described by work in Reference \cite{Barrett}. This is of particular interest for fifth force experiments due to it having the unique property of being an asymmetric geometry yet still sourcing a gravitational monopole. This novel feature could potentially be exploited when approaching Planck strength couplings for the chameleon field, where one would expect an asymmetric scalar field profile thus in principle making it possible to disentangle the two attractive forces. Collectively, these four choices explore concave and convex surfaces, and shapes that are, and aren't simply connected. 
In what follows we used a vacuum chamber of radius $R_C = 15\>\rm{cm}$, and  the width of the numerical domain was  $ 36\>\rm{cm}$. The volumes of all sources were normalized to that of a sphere with a radius of $2\>\rm{cm}$, and the density was assumed to be constant across all sources.  For the torus we take the inner radius to be fixed at $0.5\>\rm{cm}$. The ellipsoid has an ellipticity of  $\xi = 1.01$. \textcolor{blue}{ In this work we are primarily interested in the effects of changing the shape of the source mass, therefore we take the density of the source, regardless of its shape, to be the same as the density of the walls of the vacuum chamber. }
\\
Data was gathered for each of the sources by examining force scaling behaviour for certain slices through the model parameter space.  The approach involved fixing $\Lambda$ and then studying how the chameleon profile varied across a range of choices for $M$. This allowed us to cover the range of model parameters of interest, capturing how the chameleon profile for each source responds to movements in both $M$ and $\Lambda$ whilst keeping computation times down.

\begin{figure}
\begin{minipage}[c]{0.49\textwidth}
\includegraphics[width = \linewidth]{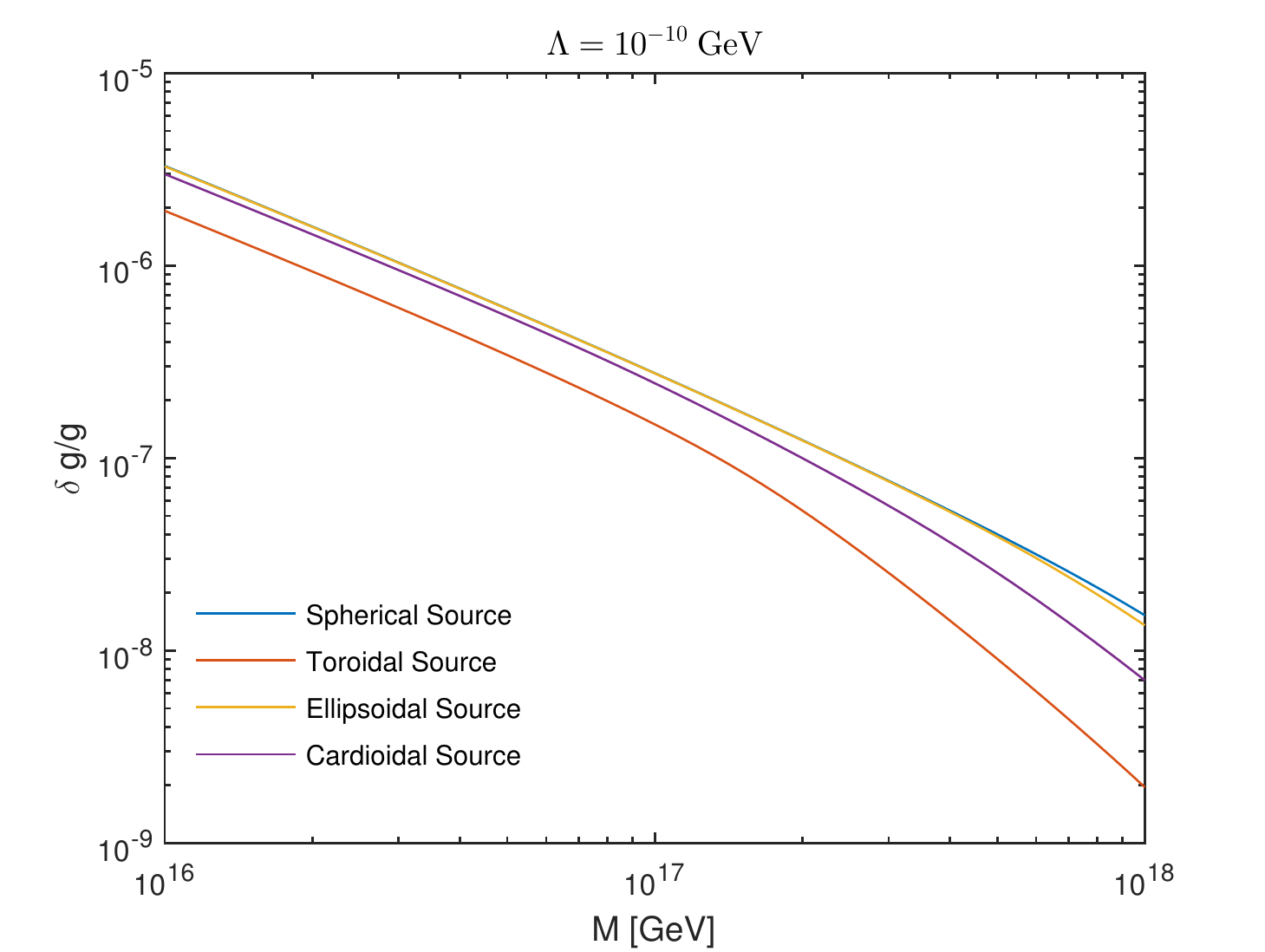}
\end{minipage}
\begin{minipage}[c]{0.49\textwidth}
\includegraphics[width = \linewidth]{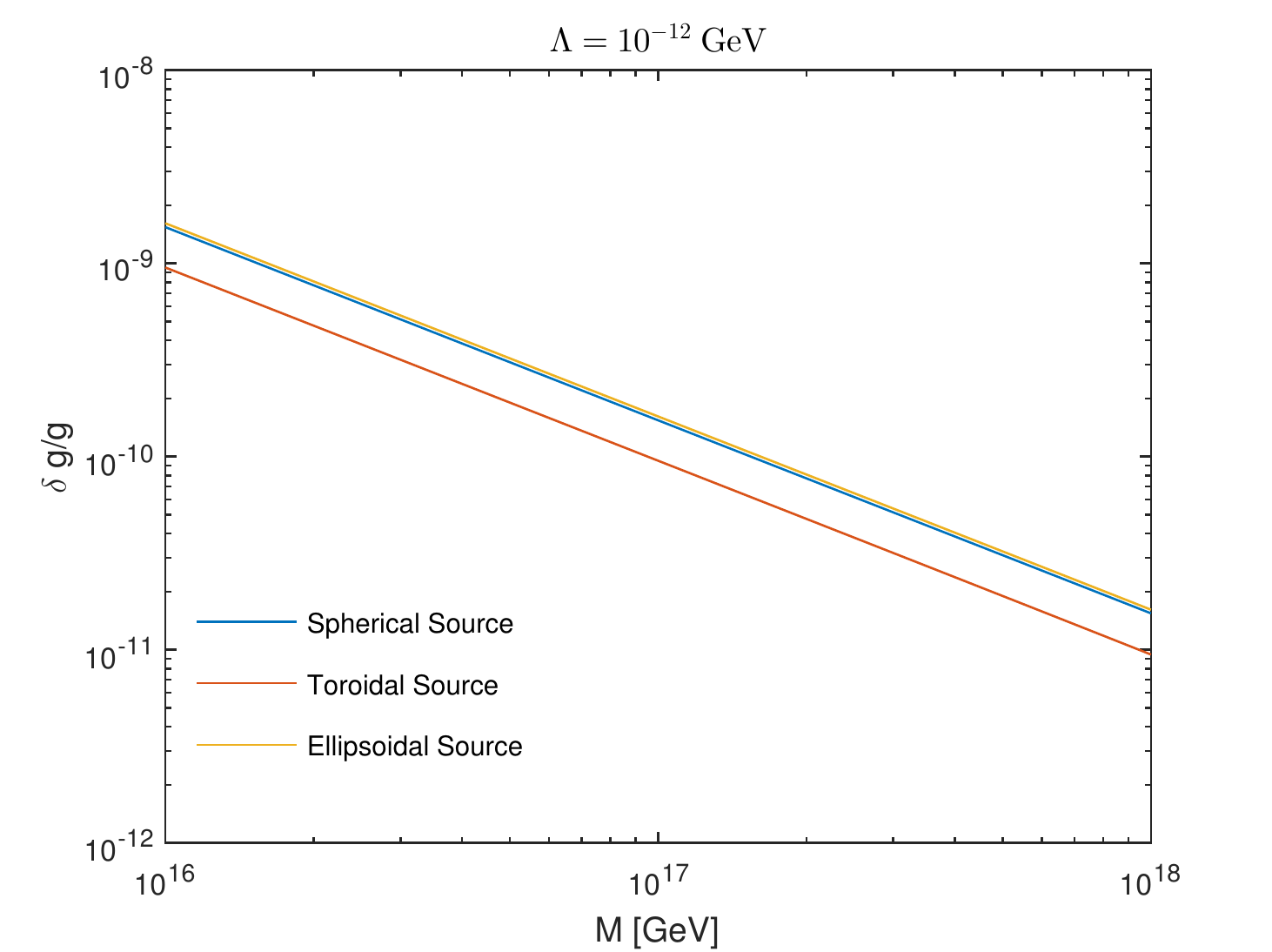}
\end{minipage}
\centering
\includegraphics[width = 0.49\linewidth]{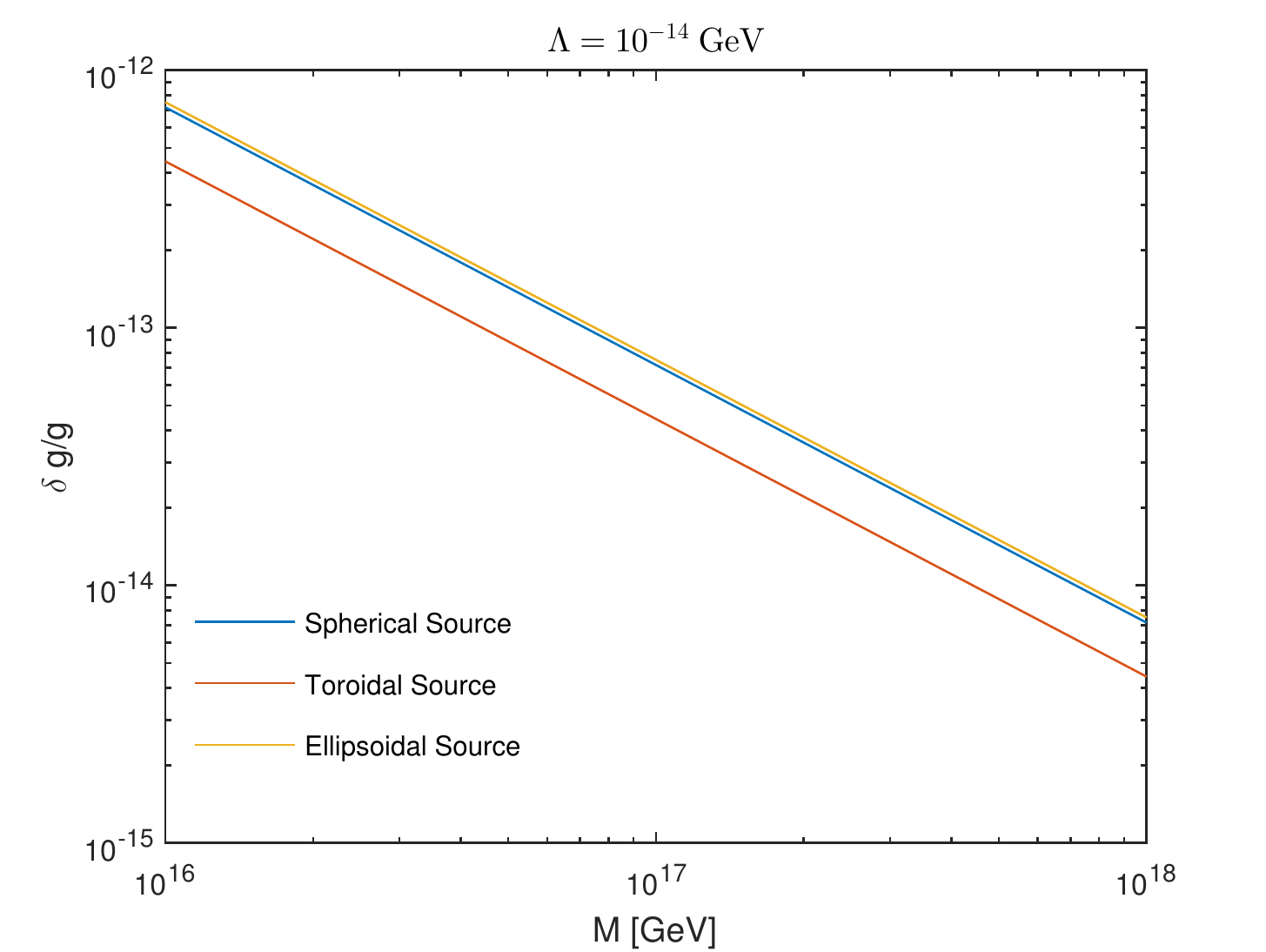}
\caption{The chameleon contribution to the  acceleration of a test particle, normalized to the Earth's gravitational field $g$, as a function of $M$, the coupling strength for three fixed values of $\Lambda$, the self interaction strength. Values of $M$ lower than $10^{16}\>\rm{GeV}$ have been omitted from the display as no additional structure is seen here. The scaling behaviour of the Cardioid source was only accessible for $\Lambda = 10^{-10}$ GeV as a result of numerical instabilities. }\label{forceScaling}
\end{figure}

Our results are displayed in Figure \ref{forceScaling}. The numerical results were verified against the known analytic solution for the spherical case \cite{Clare_ProbingDark}.
We show how the acceleration due to the chameleon experienced by a test particle varies with the coupling strength $M$ for $\Lambda= \{10^{-10}, 10^{-12}, 10^{-14}\} \mbox{ GeV}$ for the four source geometries. We plot the range $10^{16}\leq M \leq 10^{18}\>\rm{GeV}$ as no additional structure was revealed for lower values of $M$, and higher values have coupling strengths above the Planck scale. We were only able to study the cardioid for  $\Lambda = 10^{-10}\>\textrm{GeV}$, as we were unable to adequately resolve numerical instabilities for other values of the parameters. This was a result of the convexity of the object exceeding the limits of a uniform mesh. To alleviate this issue we would need to locally refine the simulation mesh in order to adequately resolve sharp variations in the density field \cite{Jamesthesis}.
Across the range of parameters considered here the toroid is found to produce the weakest acceleration. 
The ellipsoid gives the highest acceleration for lower values of $\Lambda$. 
The scaling behaviour observed here for the ellipsoid  supports our  previous analytic study \cite{Burrage:2014daa}, which showed that when the source is screened, it is less screened than the corresponding sphere.  In this case, the onset of screening is between $\Lambda = 10^{-10}\>\textrm{GeV}$ and $\Lambda = 10^{-12} \>\rm{GeV}$.
For each geometry there is a critical value of $M$ above which the thin-shell region expands sufficiently that a core region can no longer form, this is the reason behind the change in gradient of  the lines when  $\Lambda = 10^{-10}\>\textrm{GeV}$. In contrast, decreasing $\Lambda$ and keeping all other parameters fixed, drives the chameleon further into the screened regime.

For $\Lambda=10^{-10}\mbox{ GeV}$ we were able to simulate for the cardioid. Over most of the range of values of $M$ we considered the acceleration due to the cardioid closely tracks that of the  spherical and ellipsoidal solutions but falls off more for larger $M$. We conclude from this  that deviations from spherical symmetry do not always lead to an enhancement of the chameleon force. Whether an object is screened from the chameleon force, or not, depends on which point in the $M, \Lambda$ chameleon parameter space is being considered. But when two objects, of different shapes, are both screened, or both unscreened, we see that the change in shape changes the chameleon acceleration only by an overall numerical factor, and there is no futher dependence on $M$ and $\Lambda$.

\section{Shape Optimization}\label{ShapeOptimization}
The analysis of the previous section shows that  the shape of the source does affect the  acceleration due to the chameleon and that ellipsoids and spheres lead to a higher acceleration than tori, and cardioids of the same mass and density. However it is not clear from this if an ellipsoid is the optimal shape of such an experiment, or just the best of the four considered. In this section we attempt to determine an optimal source shape. 

Our strategy will be to  to construct a series of shapes using the Legendre polynomials, $P_i(\cos\theta)$,  as a basis for the shape of the surface. A range of series coefficients can be tested as a means to compare different source geometries. 
A sample space of approximately one hundred randomly generated objects will be used for analysis, and only the first four Legendre polynomials will be used in the expansion. 

In order to make direct comparisons, it is important to ensure the mass of  all the objects is the same.
For sources of  constant density, this translates to normalizing the  volume integral in spherical polar coordinates. 
\begin{eqnarray}
V &=& \int_{\Omega} r^2 \textrm{sin}(\theta)dr d\theta d\varphi\\
&=& \frac{2\pi}{3}\int_0^{\pi} [R(\textrm{cos}(\theta))]^3\textrm{sin}(\theta)d\theta
\end{eqnarray}
where expanding to fourth order in Legendre polynomials
\begin{equation}
R(\rm{cos}(\theta)) = \sum_{i=0}^3 a_i P_i\left(cos(\theta)\right)
\end{equation}
Expanding this fully we find
\begin{equation}
V = \frac{2\pi}{3}\left[2a_0^3 + 2a_0a_1^2 + \frac{6}{5}a_0a_2^2 + \frac{6}{7}a_0a_3^2 + \frac{4}{35}a_2^3 + \frac{4}{5}a_1^2a_2 + \frac{24}{105}a_2a_3^2 + \frac{36}{35}{a_1}{a_2}{a_3}\right]
\end{equation}
Requiring that the mass of this shape is the same as that of a  sphere of radius $r_{\rm{EFF}}$ requires
\begin{equation}
r_{\rm{EFF}}^3 = a_0^3 + a_0a_1^2 + \frac{3}{5}a_0a_2^2 + \frac{3}{7}a_0a_3^2 + \frac{2}{35}a_2^3 + \frac{2}{5}a_1^2a_2 + \frac{12}{105}a_2a_3^2 + \frac{18}{35}{a_1}{a_2}{a_3}
\label{eq:reff}
\end{equation}

\begin{figure}
\begin{minipage}[c]{0.329\textwidth}
\centering
\includegraphics[width=1\linewidth]{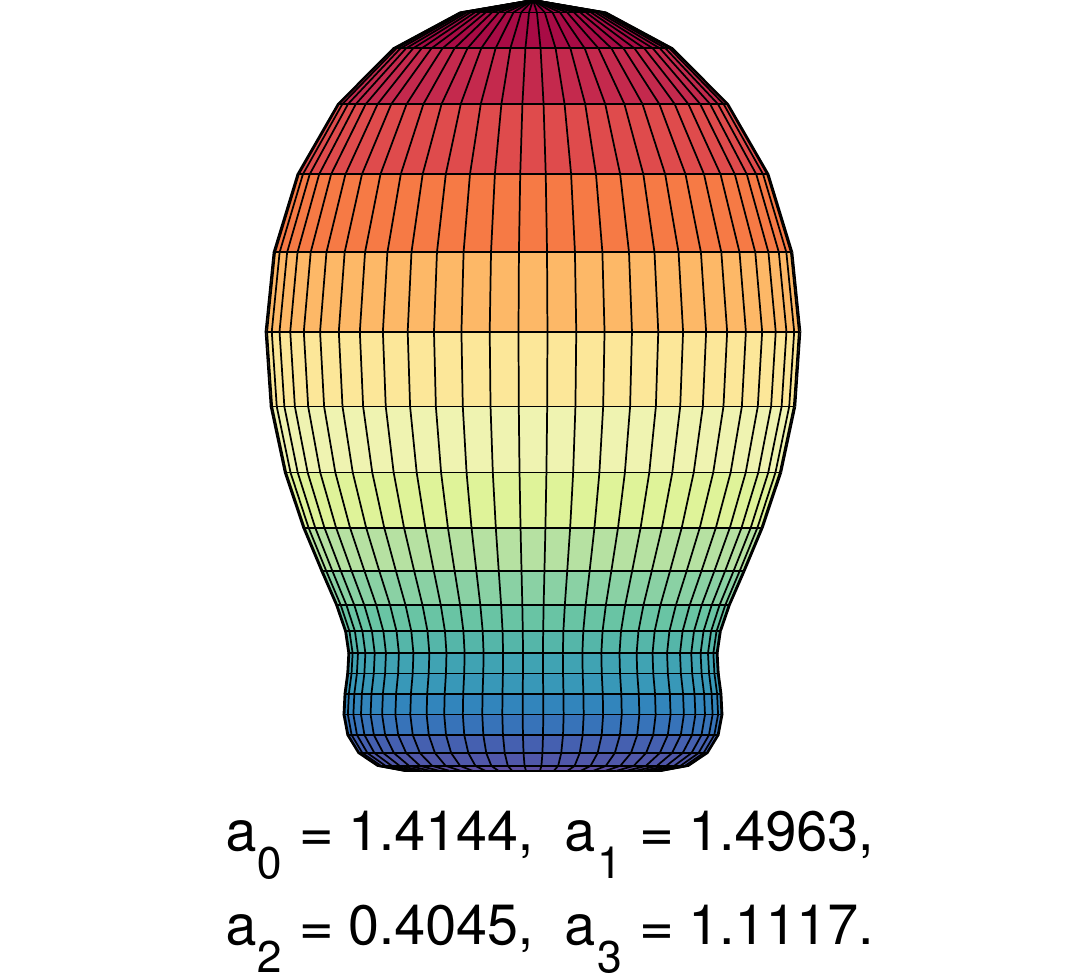}
\end{minipage}
\begin{minipage}[c]{0.329\textwidth}
\centering
\includegraphics[width=1\linewidth]{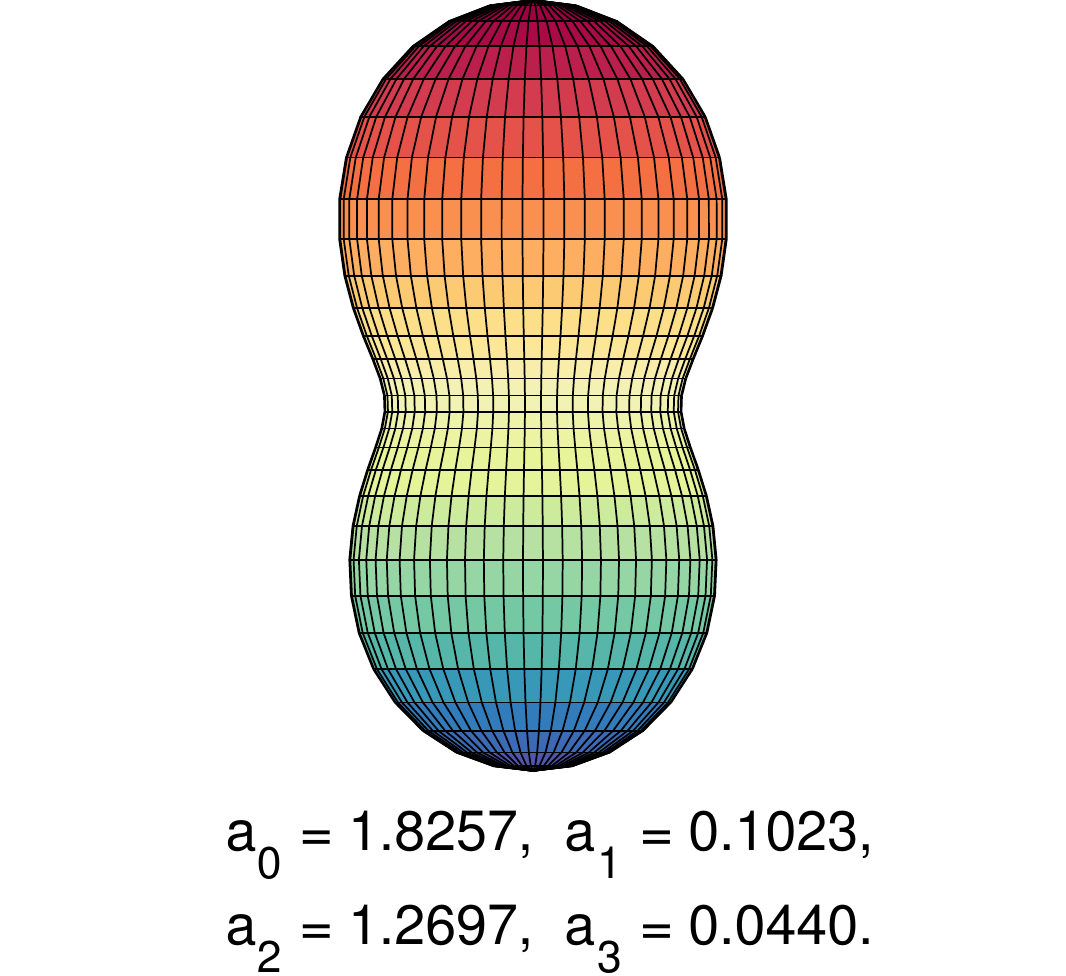}
\end{minipage}
\begin{minipage}[c]{0.329\textwidth}
\centering
\includegraphics[width=1\linewidth]{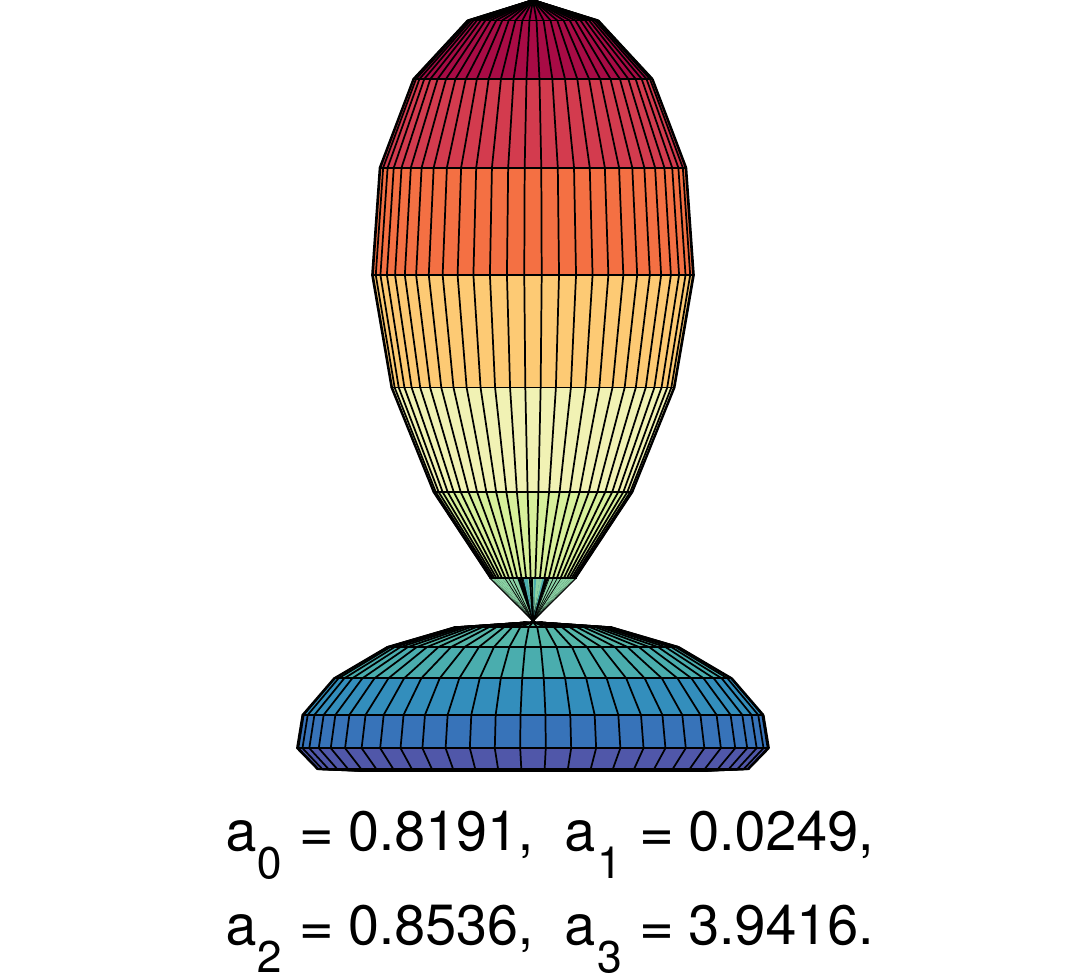}
\end{minipage}
\caption{Example geometries built using a four term Legendre Series. The coefficients of this series are  displayed beneath each source. }
\label{fig:shapes}
\end{figure}

To generate our sequence of source shapes, values of the $a_i$ are  picked sequentially from a uniform distribution, subject to the  constraint in Equation (\ref{eq:reff}), which means that values are chosen from the following ranges: 
\begin{equation}
\begin{split}
0 &\leq a_0 \leq r_{\rm{EFF}} \\ 
 0 &\leq a_1 \leq \sqrt\frac{{r_{\rm{EFF}}^3 - a_0^3}}{a_0} \\ 
0 &\leq a_3 \leq \sqrt{\frac{7}{3}\left(\frac{r_{\rm{EFF}}^3}{a_0} - a_0^2 - a_1^2\right)}
\end{split}
\end{equation}
and finally $a_2$ is determined by solving
\begin{equation}
\frac{2}{35}a_2^3 + \frac{3}{5}a_0a_2^2 + \left(\frac{2}{5}a_1^2 + \frac{12}{35}a_3^2 + \frac{36}{35}a_1a_3\right)a_2 + \left(a_0^3 + a_0a_1^2 + \frac{3}{7}a_0a_3^2 - r_{\rm{EFF}}^3\right) = 0
\end{equation}
A selection of shapes that can be built in this way are shown in Figure \ref{fig:shapes}.

For each source the finite element solver is then called to solve for the associated chameleon field within the vacuum chamber. As we are now dealing with asymmetric field profiles, we  now also employ an optimization algorithm to identify the position which would be most favourable for an experiment. To do this we  scan the  solution at a pre-set distance from the surface (taken to be 5 millimeters in this work) in order to identify the position at which the chameleon acceleration is largest.

\
\begin{table}[h!]\centering
        \pgfplotstableset{
            every head row/.style={before row=\toprule, after row=\midrule},
            every last row/.style={before row=\midrule, after row=\bottomrule}
            }
           
        \begin{filecontents*}{exportedShapeData.txt}
0.81906 0.024973 0.85368 3.9417 4.8197e-11
0.97 0.59 0.03 3.99 4.6779e-11
1.3414 0.18274 0.40532 2.8878 4.522e-11
1.4691 0.19046 0.26955 2.63 4.465e-11
2 0 0 0 1.7272e-11
\end{filecontents*}
\pgfplotstabletypeset[font =\normalsize,
            1000 sep={\,},
            columns/info/.style={
            showpos,
            column type=r,
            },
            columns/0/.style={column name={$a_0$}, fixed, precision=2,dec sep align},
            columns/1/.style={column name={$a_1$}, fixed, precision=2,dec sep align},
            columns/2/.style={column name={$a_2$}, fixed, precision=2,dec sep align},
            columns/3/.style={column name={$a_3$}, fixed, precision=2,dec sep align},
            columns/4/.style={column name={$\delta g/g$}}
            ]
            {exportedShapeData.txt}
        \caption{The change in acceleration of a test particle due to the chameleon, normalized to the acceleration due to free fall at the surface of the Earth, for a range of sources built using a 						Legendre expansion for parameters $M = 10^{18}\>\textrm{GeV}$ and $\Lambda = 10^{-12}\>\textrm{GeV}$. The first four entries correspond to the greatest accelerations returned from the sample pool. The final entry corresponds to the spherical result.}\label{proceduralTable}
\end{table}

\begin{figure}
\centering
\includegraphics[width = 0.75\linewidth]{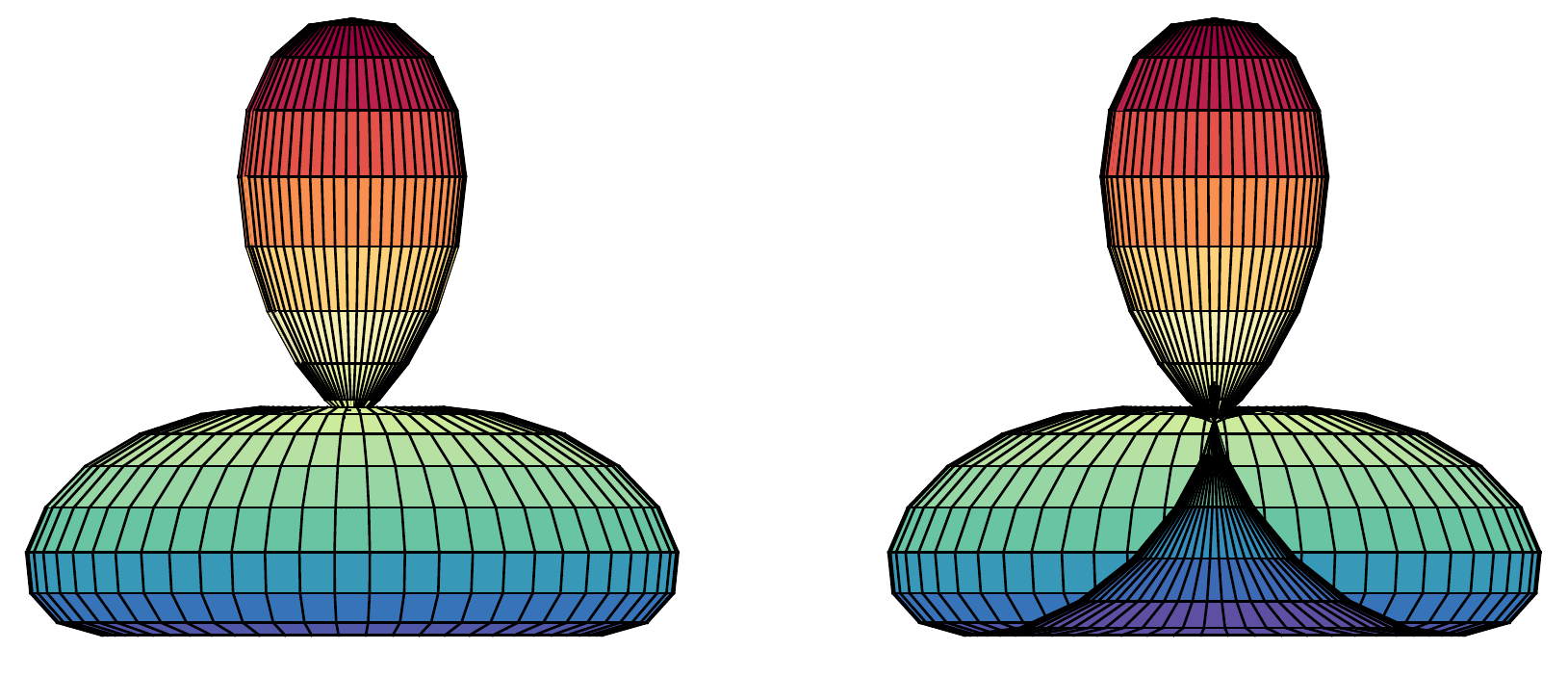}
\caption{Three dimensional display of the optimal source object extracted from the sample pool. To the left is an external view of the source whereas to the right is a view of a cross section obtained by slicing the object in half. Colours  scale with the $z$ coordinate to aid visualization.}
\label{optiSource}
\end{figure}

\begin{figure}
\centering
\begin{minipage}[c]{0.75\textwidth}
\includegraphics[width = 0.99\linewidth]{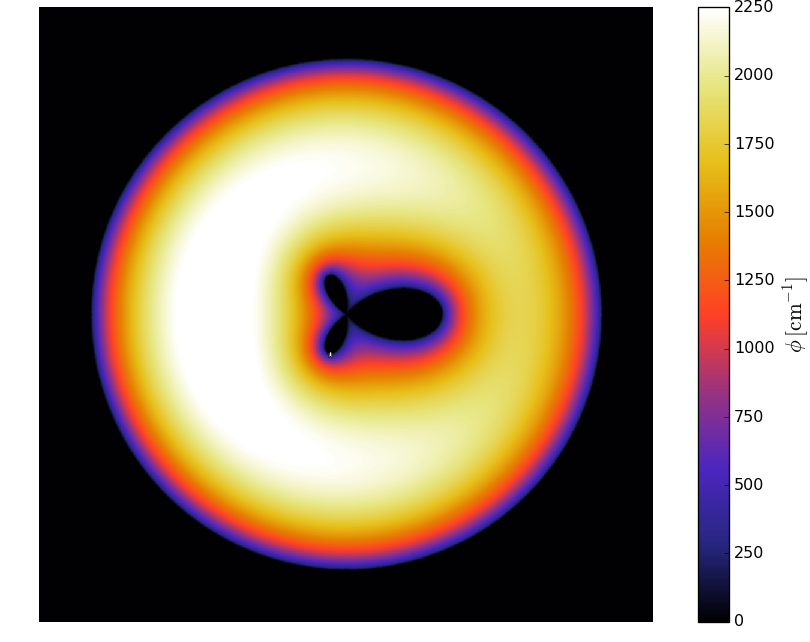}
\end{minipage}
\caption{Scalar field profile in the $(x,z)$ plane for the entry leading to the greatest acceleration
in Table \ref{proceduralTable}. As in Figure \ref{iterativeSolver}, the field ranges between the source expectation value (black) and the vacuum expectation value (white).  The optimal location for the chameleon force is signified by the small white star towards the end of the bottom left lobe. }\label{proceduralSources}
\end{figure}

The parameters which maximize the acceleration  are given  in Table \ref{proceduralTable}, with the spherical case for reference. \textcolor{blue}{For the shapes we have considered, we find a maximum increase of the chameleon acceleration by  a factor of $\sim 2.8$  over the spherical case.
The shape which gives rise to the largest acceleration} is shown in Figure \ref{optiSource} and the acceleration due to this source is maximal near the tail regions, as seen in Figure \ref{proceduralSources}. One possible reason for this could be the choke-point connecting these lobes to the main body, this could act to drive the field out of the minimum of its effective potential in the interior, essentially decoupling the adjacent regions. This will make it harder to form a  `core' region within the source, and so will reduce the screening.


These results strongly suggest that within the  screened regime the chameleon acceleration can be increased  by objects which minimize at least one of their internal dimensions.  In a different part of the chameleon parameter space, where the chameleon field can be approximated as massless outside the source object, it has previously been shown \cite{JonesSmith:2011tn}, by analogy with electrostatics, that the chameleon induced acceleration is enhanced around flattened objects. This has been dubbed the `lightning rod' effect.  In this work we have shown that thin shapes still maximise the chameleon acceleration even when the mass of the field cannot be neglected.

\section{Conclusions}
\label{sec:conc}
In this work we have implemented a finite element scheme in order to develop a better understanding of how the chameleon force responds to the geometry of the source.  We have shown that not all departures from spherical symmetry lead to an enhancement of the chameleon acceleration. By comparing the chameleon acceleration due to  spheres, ellipsoids, tori and cardioids, we have shown that which shape maximizes the chameleon acceleration depends only weakly on the values of the chameleon parameters. The only dependence is on whether the source is or is not screened.  When all objects are in the screened regime, or alternatively when all objects are unscreened, the relative chameleon acceleration due to these shapes is fixed. The onset of screening is, however, determined by the values of the chameleon parameters.

We have also determined the optimal source shape to maximize chameleon acceleration within a set of shapes where the surface is determined by a set of fourth order Legendre polynomials. This confirms our expectation that screening is decreased when objects are thinner.

These findings have implications for the design of future laboratory experiments searching for the chameleon, as modern 3D printing technology gives a huge amount of control over the shape of the source mass chosen. We have shown that, within the class of shapes we have analyzed, going to fourth order in Legendre polynomials, by optimizing the shape of the source mass the sensitivity of an experiment to the chameleon induced acceleration can be increased by a factor of up to $2.8$ compared to the spherical case.  \textcolor{blue}{It isn't clear whether this is the maximum gain that can be obtained, and more work is needed to determine whether this can be increased further.}

\section*{Acknowledgments}
We would like to thank  Jonathan Pearson for helpful discussions in the early days of this project.  CB, AM and JS are supported by the Royal Society, CB is also supported in part by a Leverhulme Trust Research Leadership Award. EJC is supported in part by STFC Consolidated Grant ST/L000393/1. Part of the work of EJC was performed at the Aspen Center for Physics, which is supported  by National Science Foundation grant PHY-1607761. 

\begin{appendix}
\section{Dimensionless form of the equations}
\label{app:dim}
In the numerical code we solve a dimensionless form of the chameleon equations of motion (\ref{fieldEquation}):
\begin{equation}
\tilde{\nabla}^2\tilde{\phi} = -\gamma_1\frac{\tilde{\Lambda}^5}{\tilde{\phi}^2} + \gamma_2\frac{\tilde{\rho}}{\tilde{M}}
\end{equation}
where 
\begin{eqnarray}
\tilde{\nabla^2} &=& \alpha_{cm}^{2}\nabla^2\\
\tilde{\phi} &=& \alpha_{cm}\phi\\
\tilde{\rho}&=& \frac{\rho}{\alpha_{g}\alpha_{cm}^{-3}}\\
\tilde{\Lambda} &=& \frac{\Lambda}{\Lambda_{DE}}\\
\tilde{M} &=& \frac{M}{M_P}
\end{eqnarray} 
and the units have been absorbed into the constants
\begin{eqnarray}
\alpha_{cm} &=& 5.1\times 10^{13}\>\rm{GeV}^{-1}\\
\alpha_{g} &=& 5.62\times 10^{23}\>\rm{GeV}\\
\Lambda_{DE} &=& 2.4 \times 10^{-12}\>\rm{GeV}\\
\gamma_1&=& (\Lambda_{DE} \alpha_{cm})^5\\
\gamma_2 &=& \frac{\alpha_{g}}{M_P}
\end{eqnarray}
The force that the chameleon exerts on a unit test mass is given by
\begin{equation}
\vec{F}=-\frac{\vec{\nabla}\phi}{M}
\end{equation}
This is obviously equal to the acceleration experienced by the test mass.  When the time evolution of the system is required we solve 
using a  traditional leapfrog scheme. If  $\dot{\phi} = \pi$, then the dimensionless form of this equation is
\begin{equation}
\tilde{\dot{\pi}} = -\frac{\gamma_3}{\tilde{M}}\tilde{\nabla}\tilde{\phi} 
\end{equation}
where
\begin{eqnarray}
\alpha_s &=& 1.51925\times 10^{24}\>\rm{GeV}^{-1}\\
\gamma_3& =& \frac{\alpha_{\rm{s}}^2}{\alpha_{\rm{cm}} M_{\rm{PL}}}
\end{eqnarray}

\end{appendix}

\end{document}